\documentclass[smallextended]{svjour3}       
\usepackage{rotating}
\usepackage{longtable,geometry}
\usepackage[utf8]{inputenc}
\usepackage[babel]{csquotes}
\usepackage[usenames,dvipsnames,svgnames,table]{xcolor}
\usepackage{float}

\usepackage{amsfonts}
\usepackage{amsmath, amssymb}
\usepackage{setspace}
\usepackage{vmargin}
\usepackage{graphicx,psfrag}

\usepackage{rotating}

\def\nuc{\text{nuc}}
\def\el{\text{el}}
\def\hrho{\hat{\rho}}

\pdfmapfile{=mtpro2.map}

\newcommand{\cm}{cm$^{-1} \ $}

\newcommand{\sch}{Schr\"{o}dinger\ }

\newcommand{\JCPformat}[4]{{#1} {\bf #2}, {#3} ({#4}).}

\newcommand{\Refs}[4]{\JCPformat{#1}{#2}{#3}{#4}}

\newcommand{\jcpp}[3]{\Refs{J. Chem. Phys.}{#1}{#2}{#3}}

\newcommand{\cpl}[3]{\Refs{Chem. Phys. Lett.}{#1}{#2}{#3}}
\newcommand{\chemphys}[3]{\Refs{Chem. Phys.}{#1}{#2}{#3}}
\newcommand{\tca}[3]{\Refs{Theor. Chim. Acta.}{#1}{#2}{#3}}

\newcommand{\jmathchem}[3]{\Refs{J. Math. Chem.}{#1}{#2}{#3}}
\newcommand{\jmathphys}[3]{\Refs{J. Math. Phys.}{#1}{#2}{#3}}

\newcommand{\molphys}[3]{\Refs{Mol. Phys.}{#1}{#2}{#3}}

\newcommand{\ijqc}[3]{\Refs{Int. J. Quantum Chem.}{#1}{#2}{#3}}

\newcommand{\physrev}[3]{\Refs{Phys. Rev.}{#1}{#2}{#3}}
\newcommand{\physrevlett}[3]{\Refs{Phys. Rev. Lett.}{#1}{#2}{#3}}

\newcommand{\jphys}[3]{\Refs{J. Phys.}{#1}{#2}{#3}}
\newcommand{\revmodphys}[3]{\Refs{Rev. Mod. Phys.}{#1}{#2}{#3}}

\DeclareFontFamily{U}{mathb}{\hyphenchar\font45}
\DeclareFontShape{U}{mathb}{m}{n}{
      <5> <6> <7> <8> <9> <10> gen * mathb
      <10.95> mathb10 <12> <14.4> <17.28> <20.74> <24.88> mathb12
      }{}
\DeclareSymbolFont{mathb}{U}{mathb}{m}{n}
\DeclareMathSymbol{\dlsh}{3}{mathb}{"EA}

\begin{document}


\title{Electrons as an environment for nuclei within molecules: a quantitative assessment of their contribution to a classical-like molecular structure}

\author{Patrick Cassam-Chena\"\i \and Edit M\'atyus}

\institute{F. Author \at
Universit\'e C\^ote d'Azur, CNRS, LJAD, UMR 7351, 06100 Nice, France.
\email{cassam@unice.fr}
   \and
           S. Author \at
ELTE, Eötvös Loránd University, Pázmány Péter sétány 1/A, 1117 Budapest, Hungary.
\email{matyus@chem.elte.hu}}

\date{Received: date / Accepted: date}

\maketitle

\begin{abstract}
Molecular structure is often considered as emerging from the decoherence effect of the environment. Electrons are part of the environment of the nuclei in a molecule. In this work, their contribution to the classical-like geometrical relationships often observed between nuclei in molecular systems  is investigated. Reduced density matrix (RDM) elements are evaluated from electron-nucleus wave functions. The computational results show that the electrons play a role in the localization of the nuclei around specific geometries. Although the electronic environment alone cannot explain molecular symmetry-broken isomers, it can contribute to their dynamical stability by reducing off-diagonal RDM elements.
\end{abstract}
\maketitle


Suggested running head:
Decoherence within molecules

Correspondance can be addressed to P. Cassam-Chena\"{\i},\\
cassam@unice.fr,\\
tel.: +33 4 92 07 62 60,\\
fax:  +33 4 93 51 79 74\\

\newpage

\section{Introduction}

The usual approach for reconstructing or recognizing molecular structural elements from a  wave function 
follow the observation of Claverie and Diner \cite{Claverie1980}, that classical structures can be identified with nuclear
configurations for which  appropriately defined density functions have maxima. 
Within such a view, based on Born's probabilistic interpretation of the square modulus of the wave function, 
molecules do exhibit clear structural features as demonstrated by accurate calculation of their full, \emph{i.e.,} ``all-particle'' wave functions,
in the sense that inter-nuclei geometrical parameter densities are peaked at definite values \cite{MaHuMuRe11a,MaHuMuRe11b}. However, when identical nuclei are present, the averaging over the permutational symmetry group, spoils the relevance of these features for retrieving a classical molecular structure, as noticed in many instances \cite{Sutcliffe05}.

Decoherence effects by the environment \cite{Joos03} are often invoked to explain why molecule 
 behave as near classical objects with structural features related to those maximal density configurations,  
 that chemists can use without having usually to worry about any quantum mechanical
interference phenomena. In this article, we focus on these ``Classical-like'' structure (a precise definition will be proposed in section  \ref{Classical-like}), and refer the reader interested in quantum nuclear effects in molecules to a few recent reviews \cite{Cronin2009,Hornberger2012,Koch2019,Kastner2020}.
The environment of a molecular system has undoubtly some decoherence effects and something
to do with the localization of the system in a state with ``quasi-classical'' characteristics.
But when we are thinking about molecules, it is hard to imagine a completely generic environment. 
How to formulate in mathematical expressions such a general i.e non specific environment? 
There have been proposals to consider the photon vacuum field as an ubiquitously present environment, 
responsible for the stability of isolated, chiral molecules  \cite{Pfeifer81,Primas80,Primas81-lnc}. However,
it has been shown that the proposed mechanism was only valid at zero temperature \cite{Wightman95}.
In the early years of the development of decoherence ideas in connection with the molecular structure
problem, Claverie and Jona-Lasinio  \cite{Claverie1986,Jona-Lasiano86} used external random noise to simulate 
localization in a double potential well (which is a typical toy model for  the ammonia
``umbrella'' inversion and parity-breaking in chiral molecules \cite{Hund1927,Quack2014}). However, the reaction field mechanism of these authors
is a collective effect, difficult to use convincingly for quasi-isolated molecules, as can be found in astrophysical conditions, 
where densities of one molecule per cubic centimeters or less, are observed.
Davies argued that, for a collection of identical molecules (at least two), there exists metastable approximate
eigenstates in the form of a tensor product of one and the same molecular state, which are both close to the genuine eigenstates 
of the whole collection of molecules, and symmetry-broken with respect to the individual molecule's symmetry group~\cite{Davies95}.
However, it remains to justify why the whole system would be in such an approximate product state rather than in a true eigenstate.
The spin-boson model can encompass a variety of environments, such as the electromagnetic radiation field, 
as long as they can be represented by a set of model harmonic oscillators within some simplifying hypotheses \cite{Amann91}.
Hornberger and co-workers simulated the stabilization of chiral molecules upon collisions \cite{Hornberger2007,Trost2009,Busse2010}
and studied the orientational decoherence of molecules and nanoparticles \cite{Stickler2018a,Stickler2018b}.
Recent realistic decoherence simulations demonstrate that different environment models 
have different decoherence properties that affect different degrees of freedom differently \cite{Zhong2016}. 
These simulations are very interesting, because the systematic
and accurate calculation of the decoherence times for a variety of molecular 
processes, in interaction with a series of ``standard'' environment models, 
could be useful for controlling decoherence in real systems 
and designing better quantum computers (using molecular qubits).

In the present work, we prefer to confine ourself to the sole molecular system: we are
seeking the furthermost point, one can reach in resolving the molecular structure
conundrum~\cite{Weininger84,Woolley85,Lowdin1989}, without explicitly considering any specific kind of environment. By molecular structure, 
we understand the relative localization of the nuclei in the three dimensional space. However, a
molecule consist of not only nuclei but also of electrons. So, it is appropriate to ask to
which extent the electrons play a role in the localization of the nuclei by their continuous monitoring.
The idea is that nuclei are constantly ``measured'' by electrons through their Coulomb interactions. 
Models such as that of Ref.~\cite{Hu2018},  found an initial decoherence time due 
to electrons of the order of a few femtoseconds, that is much faster than the characteristic time scale for nuclear ro-vibrational motion.
However, using only two electronic basis functions obtained as eigenfunctions of a clamped nuclei Hamiltonian, 
as  in \cite{Hu2018}, would be questionable for a real molecular system. 
Time propagation of an initial pure state would lead to the ground state of the system, 
as in Monte-Carlo simulation, and such a state should be decomposed on a complete, 
infinite dimensional basis set of electronic states.
The purpose of this work is to study the localization and decoherence effects of the electrons on the nuclei,
by using accurate electron-nucleus, molecular wave functions. By decoherence, we mean essentially here, the suppression of off-diagonal matrix elements of the reduced density matrix for the nuclei alone.

The article is organized as follows:
In the next section, we introduce the concept of ``pointer states'', define what we mean by a ``classical-like'' molecular structure and present the notion of ``purity'' of the reduced density matrix (RDM), 
for the nuclei of a molecule, the electrons, considered as the environment of the latter, being traced out. 
Since, this electronic environment corresponds only to a finite set 
of degrees of freedom, we cannot expect superselection rules to emerge, 
but we are curious about what kind of conclusions can be drawn within this setup. 
This is investigated in the third section,
before concluding in the last section.


\section{Theoretical tools to quantify the classicality of molecular structure \label{sec:moldec}}

Let us first define in a very pedestrian way, the basic theoretical tools, we will rely on in the rest of the paper.

\subsection{Pointer states}

When measuring a property of a quantum system, the needle of an (idealized) measuring device points to one of the possible outcome values.
In a satisfactory theory of quantum measurement, an experimental setup, although macroscopic, should be amenable to a quantum treatment. 
Hence, the idea to associate a quantum state to every position of the needle. These states were termed ``pointer state'' by Zurek \cite{Zurek1981}, and their apparent classical behaviour was assumed to be due to the decoherence effect of the environment. 

At present, in decoherence theory, the concept of ``pointer states'' has been extended to a wider context, where there is not necessarily a \textit{bona fide} experimental setup.  The environment of a quantum system is assumed for all practical purposes, to break the unitary invariance of the quantum mechanical representation of the system, by selecting a special basis in which the
``coherences'', \textit{i.e.} the non diagonal elements of the density matrix, decrease exponentially with time. The ``pointer states'' are defined as the pure states belonging to the basis set selected by the environment, the latter constantly destructing their superposition.

There is no general theory to determine the pointer states of a quantum system in a given environment. 
For each  microscopic  environment modelling, one has to tackle the task of finding the proper pointer states \cite{Busse2010,Zurek1981}.
 However, in many cases,  such as macroscopic objects which appear perfectly localized in space, the representation selected by the environment
 is the so-called ``direct representation'', the pointer states corresponding to Dirac distributions in configuration space.

\subsection{Reduced density and transition operator matrices for nuclear degrees of freedom (DOFs)}
\subsubsection{Pure states}
Let $|\Psi\rangle$ be a molecular, normalized wave function in Dirac ket notation. The associated (pure state) density operator, $|\Psi\rangle\langle\Psi|$, will be denoted as  $\hrho$. The representation selected by an environment being often the ``direct representation'', let us consider it first to express the density operator matrix.  In the direct representation, denoting collectively by $\mathbf{r}$ the electronic DOFs coordinates and by 
$\mathbf{R}$ the nuclear ones, we have,
\begin{eqnarray}
 \hrho&=\int d\mathbf{r}\ d\mathbf{R}\ |\mathbf{r}\ \mathbf{R}\rangle\langle\mathbf{r}\ \mathbf{R}|\cdot|\Psi\rangle\langle\Psi|\cdot\int  d\mathbf{r'}\ d\mathbf{R'}\ |\mathbf{r'}\ \mathbf{R'}\rangle\langle\mathbf{r'}\ \mathbf{R'}|&\nonumber\\
  &=\int d\mathbf{r}\ d\mathbf{R}\ d\mathbf{r'}\ d\mathbf{R'}\ |\mathbf{r}\ \mathbf{R}\rangle\langle\mathbf{r}\ \mathbf{R}|\Psi\rangle\langle\Psi|\mathbf{r'}\ \mathbf{R'}\rangle\langle\mathbf{r'}\ \mathbf{R'}|&\nonumber\\
   &=\int d\mathbf{r}\ d\mathbf{R}\ d\mathbf{r'}\ d\mathbf{R'}\ \Psi^*(\mathbf{r'},\mathbf{R'})\Psi(\mathbf{r},\mathbf{R})\ |\mathbf{r}\ \mathbf{R}\rangle\langle\mathbf{r'}\ \mathbf{R'}|.&
    \label{rdm}
\end{eqnarray} 
To study the nuclear structure, we integrate out the electronic
degrees of freedom that are considered as the environment for the nuclei.
The resulting reduced density matrix operator for the nuclear motion is
\begin{eqnarray}
 \hat{\rho}_\nuc
  &=
  \text{Tr}_\el\ [\hat{\rho} ]
  \nonumber \\
&=\int d\mathbf{r''}\ \langle\mathbf{r''}|\Psi\rangle\langle\Psi|\mathbf{r''}\rangle&\nonumber\\
   &=\int d\mathbf{r''}\ d\mathbf{r}\ d\mathbf{R}\ d\mathbf{r'}\ d\mathbf{R'}\ \Psi^*(\mathbf{r'},\mathbf{R'})\Psi(\mathbf{r},\mathbf{R})\  \langle\mathbf{r''}|\mathbf{r}\ \mathbf{R}\rangle\langle\mathbf{r'}\ \mathbf{R'}|\mathbf{r''}\rangle&\nonumber\\
      &=\int d\mathbf{r''}\ d\mathbf{r}\ d\mathbf{R}\ d\mathbf{r'}\ d\mathbf{R'}\ \Psi^*(\mathbf{r'},\mathbf{R'})\Psi(\mathbf{r},\mathbf{R})\ \delta_{r''r'}\delta_{r''r} |\mathbf{R}\rangle\langle\mathbf{R'}|&\nonumber\\
           &=\int d\mathbf{R}\ d\mathbf{R'} \left(  \int d\mathbf{r''}\ \Psi^*(\mathbf{r''},\mathbf{R'})\Psi(\mathbf{r''},\mathbf{R})\right)  |\mathbf{R}\rangle\langle\mathbf{R'}|,&
            \label{rdm_nuc}
\end{eqnarray} 
so that, 
\begin{eqnarray}
\rho_\nuc(\mathbf{R},\mathbf{R'}):= \langle\mathbf{R}|\hat{\rho}_\nuc|\mathbf{R'}\rangle= \int d\mathbf{r''}\ \Psi^*(\mathbf{r''},\mathbf{R'})\Psi(\mathbf{r''},\mathbf{R})
 \label{rdm_nuc_el}
\end{eqnarray}


If the wave function assumes a Born--Oppenheimer (BO) form, $\Psi_{BO}(\mathbf{r},\mathbf{R})=\Psi_{e}(\mathbf{r},\mathbf{R})\Psi_{N}(\mathbf{R})$, then Eq. (\ref{rdm_nuc_el}) becomes 
\begin{eqnarray}
 \langle\mathbf{R}|\hat{\rho}_\nuc|\mathbf{R'}\rangle &=\left(  \int d\mathbf{r''}\ \Psi_e^*(\mathbf{r''},\mathbf{R'})\Psi_e(\mathbf{r''},\mathbf{R})\right) \Psi^*_{N}(\mathbf{R'})\Psi_{N}(\mathbf{R})\  ,&
  \label{rdm_nuc_BO_el}
\end{eqnarray} 
that is to say, the interference amplitude between pointer states $|\mathbf{R}\rangle$ and $|\mathbf{R'}\rangle$ for the nuclear system depends upon the overlap of the BO electronic functions, $\int d\mathbf{r''}\ \Psi_e^*(\mathbf{r''},\mathbf{R'})\Psi_e(\mathbf{r''},\mathbf{R})$.\\
In the case of Refs. \cite{Hu2018,Cassam15-pra,Cassam17-tca} an all-particle wave function is written in a tensor product basis as, $\Psi_{TPB}(\mathbf{r},\mathbf{R})=\sum\limits_{i,I}\lambda_{i,I}\Psi_{e}^i(\mathbf{r})\Psi_{N}^I(\mathbf{R})$, (all basis sets are taken orthonormal). Then  Eq. (\ref{rdm_nuc_el}) reads 
\begin{eqnarray}
 \langle\mathbf{R}|\hat{\rho}_\nuc|\mathbf{R'}\rangle &=\sum\limits_{i,I,j,J}\lambda_{i,I}^*\lambda_{j,J}\left(  \int d\mathbf{r''}\ \Psi_e^{*i}(\mathbf{r''})\Psi_e^j(\mathbf{r''})\right) \Psi^{*I}_{N}(\mathbf{R'})\Psi_{N}^J(\mathbf{R})\  &\nonumber\\
 &= \sum\limits_{i,I,j,J}\lambda_{i,I}^*\lambda_{j,J}\delta_{i,j} \Psi^{*I}_{N}(\mathbf{R'})\Psi_{N}^J(\mathbf{R})\  &\nonumber\\
  &= \sum\limits_{I,J}\left( \sum\limits_{i}\lambda_{i,I}^*\lambda_{i,J} \right) \Psi^{*I}_{N}(\mathbf{R'})\Psi_{N}^J(\mathbf{R})\  .&
  \label{rdm_nuc_TPB_el}
\end{eqnarray} 
This shows that all electronic functions contribute to the interference amplitude between pointer states $|\mathbf{R}\rangle$ and $|\mathbf{R'}\rangle$ through $\sum\limits_{i}\lambda_{i,I}^*\lambda_{i,J}$, which is nothing but the reduced density matrix element in the $\left(\Psi^{I}_{N}\right)_I$ basis:
\begin{eqnarray}
 \langle \Psi^{J}_{N}|\hat{\rho}_\nuc|\Psi^{I}_{N}\rangle &= \sum\limits_{i}\lambda_{i,I}^*\lambda_{i,J},&
\end{eqnarray} 
as can be seen by comparing Eq. (\ref{rdm_nuc_TPB_el}) with the change of representation formula:
\begin{eqnarray}
\langle\mathbf{R}|\hat{\rho}_\nuc|\mathbf{R'}\rangle &= \sum\limits_{I,J} \langle\mathbf{R}|\Psi^{J}_{N}\rangle \langle \Psi^{J}_{N}|\hat{\rho}_\nuc|\Psi^{I}_{N}\rangle \langle\Psi_{N}^I|\mathbf{R'}\rangle\ .&
\end{eqnarray} 
We note in passing, that, would the pointer state basis be a general one, such as $\mathbf{B}_{nuc}:=\left(\Psi^{I}_{N}\right)_I$, instead of $\left(|\mathbf{R}\rangle\right)_{\mathbf{R}}$, the corresponding reduced density matrix elements could be easily derived owing to this transformation.

\textit{Remark 1:}
One can define reduced transition matrices (RTM) in a similar fashion. Let $|\Psi_1\rangle\langle\Psi_2|$ be the transition operator from molecular state $\Psi_2$ to $\Psi_1$, the reduced  transition matrix elements in the direct representation are,
\begin{eqnarray}
 \langle\mathbf{R}|RTM_{nuc}|\mathbf{R'}\rangle= \int d\mathbf{r''}\ \Psi_2^*(\mathbf{r''},\mathbf{R'})\Psi_1(\mathbf{r''},\mathbf{R})\ .
 \label{rtm_nuc_el}
\end{eqnarray}
When $\Psi_1$ and $\Psi_2$ are decomposed on a tensor product basis set, $\Psi_{1}(\mathbf{r},\mathbf{R})=\sum\limits_{i,I}\lambda^1_{i,I}\Psi_{e}^i(\mathbf{r})\Psi_{N}^I(\mathbf{R})$ and $\Psi_{2}(\mathbf{r},\mathbf{R})=\sum\limits_{i,I}\lambda^2_{i,I}\Psi_{e}^i(\mathbf{r})\Psi_{N}^I(\mathbf{R})$, one obtains,
\begin{eqnarray}
 \langle\mathbf{R}|RTM_{nuc}|\mathbf{R'}\rangle
  &= \sum\limits_{I,J}\left( \sum\limits_{i}\lambda_{i,I}^{2\ ^*}\lambda_{j,J}^1\right) \Psi^{*I}_{N}(\mathbf{R'})\Psi_{N}^J(\mathbf{R})\  .&
  \label{rtm_nuc_TPB_el}
\end{eqnarray} 
In the case of two BO wave functions, $\Psi_{1}(\mathbf{r},\mathbf{R})=\Psi^1_{e}(\mathbf{r},\mathbf{R})\Psi^1_{N}(\mathbf{R})$ and $\Psi_{2}(\mathbf{r},\mathbf{R})=\Psi^2_{e}(\mathbf{r},\mathbf{R})\Psi^2_{N}(\mathbf{R})$ one has more simply,
\begin{eqnarray}
 \langle\mathbf{R}|RTM_{nuc}|\mathbf{R'}\rangle &=\left(  \int d\mathbf{r''}\ \Psi^{2\ ^*}_e(\mathbf{r''},\mathbf{R'})\Psi^1_e(\mathbf{r''},\mathbf{R})\right) \Psi^{2\ ^*}_{N}(\mathbf{R'})\Psi^1_{N}(\mathbf{R})\  .&
  \label{rtm_nuc_BO_el}
\end{eqnarray}
Time dependence has only been implicit, so far. If $\Psi_1$ and $\Psi_2$, are stationary eigenstates of the total Hamiltonian associated to eigenvalues $E_1$ and $E_2$, the RTM will oscillate as $e^{-\frac{i(E_1-E_2)\cdot t}{\hbar}}$, while the RDM of a stationary state will be time independent.

\subsubsection{Ensemble states \label{ens_state}}
In this work, we will be dealing with eigensolutions of the time-independent \sch equation. When such a solution is  degenerate, a density matrix must be used.  When no particular component has been selected by the experimental setting (in the case of an experiment) or by the natural physico-chemical conditions (in the case of observations of a remote medium), it is common practice to represent a degenerate eigenstate by a generic density operator that is a convex combination of degenerate pure states having all the same probability. (This amounts to assume a Boltzmann distribution, since all pure states have the same energy). More explicitly,  the density operator is taken to be the sum with equal weights of the pure state density operators of an orthonormal set of degenerate eigenfunctions $(|\Psi_i\rangle)_{i\in\{1,\ldots ,n\}}$,
\begin{eqnarray}
 \hrho&=\frac{1}{n}\sum\limits_{i=1}^n\ |\Psi_i\rangle\langle\Psi_i|,
    \label{ens_rdm}
\end{eqnarray} 
the normalization factor, $\frac{1}{n}$, insuring that $Tr[\hrho]=1$. 
This permits to  treat all components on an equal footing, and to retrieve correct line strengths in spectroscopy, for instance.

Such an operator can be reduced by tracing out electronic DOFs as for a pure state density operator
\begin{eqnarray}
 \hat{\rho}_\nuc =  \text{Tr}_\el\ [\hat{\rho} ]  
           = \frac{1}{n}\sum\limits_{i=1}^n\  \hat{\rho}_{\nuc}^i ,
            \label{ens_rdm_nuc}
\end{eqnarray} 
where,
\begin{eqnarray}
 \hat{\rho}_{\nuc}^i = \int d\mathbf{R}\ d\mathbf{R'} \left(  \int d\mathbf{r''}\ \Psi_i^*(\mathbf{r''},\mathbf{R'})\Psi_i(\mathbf{r''},\mathbf{R})\right)  |\mathbf{R}\rangle\langle\mathbf{R'}|.
            \label{rdm_nuc_i}
\end{eqnarray} 
It is of interest to distinguish two extreme cases. In the first case, the reduced pure state operators, $ \hat{\rho}_{\nuc}^i$, are all equal to one and the same operator denoted by, say, $ \hat{\rho}_{\nuc}^0$:
\begin{eqnarray}
\forall i, \quad  \hat{\rho}_{\nuc}^i = \hat{\rho}_{\nuc}^0.
\label{elec-deg}
\end{eqnarray} 
This can be the case for an electron spin multiplet, if the electronic spin is not coupled  to the nuclear (whether spatial or spin) angular momenta (or if the coupling is neglected). That is to say, the initial $|\Psi_i\rangle$'s are orthogonal electron spin components, and after tracing over the electron spin DOFs, they produce identical functions of the nuclear variables. Then, in such a limit case, the nuclear RDM has a single term,  $\hat{\rho}_{\nuc}=\hat{\rho}_{\nuc}^0$,
as for a non-degenerate molecular state.

In the second case, the reduced pure state operators, $ \hat{\rho}_{\nuc}^i$, are still orthogonal, in the sense that,
\begin{eqnarray}
\forall i\neq j, \quad \text{Tr}_\nuc[ \hat{\rho}_{\nuc}^i \hat{\rho}_{\nuc}^j ]=0,
\label{ens_orth}
\end{eqnarray} 
so that 
\begin{eqnarray}
\text{Tr}_\nuc[\hat{\rho}_{\nuc}^2] = \frac{1}{n^2} \sum\limits_{i=1}^n \text{Tr}_\nuc[\left(\hat{\rho}_{\nuc}^i\right)^2] \leq  \frac{1}{n^2} \sum\limits_{i=1}^n \text{Tr}_\nuc[\left(\hat{\rho}_{\nuc}^i\right)] \leq \frac{1}{n} .
\label{ens_ineq}
\end{eqnarray} 
This last inequality implies that $\hat{\rho}_{\nuc}$ is still an ensemble state, sum of no less than $n$ pure state operators.
This will be the case of a nuclear spin multiplet, not coupled to electronic (whether spatial or spin) angular momenta, for which the tracing out of electronic DOFs will not alter orthogonality. \\

In the general case, the expression of the reduced density operator $ \hat{\rho}_{\nuc}$ can be simplified into a linear combination of a number $k< n$ of reduced density operators derived from molecular pure states, if and only if linear dependencies  between the  $\hat{\rho}_{\nuc}^i$'s occur. 

\textit{Remark 2:}
The $\Psi_i$ being all stationary eigenstates of the total Hamiltonian associated to the same eigenvalue, the ensemble state density operator $\hat{\rho}$ and its reduced density operator $\hat{\rho}_\nuc$ will be time independent.

\textit{Remark 3:} For a molecule in a general environment, the definition of the reduced density matrix operator for the nuclear motion, $\hat{\rho}_\nuc$, is formally identical. We only need to start from the total wave function of the molecule plus its environment, and to integrate out both the electronic and environmental DOFs.

\subsection{Classical-like molecular structure \label{Classical-like}}
We will say that a property of a quantum system is ``classical-like'', to distinguish it from ``truly quantum''
or from ``chaotic'', if the outcomes of its measurement have a narrow distribution, compatible with what one would
expect for a plausible experimental uncertainty distribution of a classical property measurement.

This implies two constraints on the reduced density operator of the system after tracing out the environment degrees-of-freedom:
(i) In the ``pointer state'' basis representation where the RDM is diagonal, all the significant eigenvalues (which give the probabilities
to obtain the corresponding eigenstates) must correspond to eigenstates which give expectation values for the property falling within a
narrow distribution
(ii) Decoherence must rapidly lead to the decay of any superposition of pointer states (related to the environment monitoring) back to the mixture of (i), after a perturbation of the system such as the measurement of the property of interest (which would project the system to a pointer state associated to the property measuring device, so \textit{a priori} to a superposition of environment-selected pointer states).
 
\subsection{Purity}

If there are fewer pointer states with a significant probability, then it is easier to fulfill condition (i). The limit case, where one pointer state has probability close to one, and therefore all the others have a probability close to zero, is the most favorable to deal with, because then one can assume that the environment will select this pointer state and one has just to verify that the property has a narrow distribution of possible measurement outcomes for that pointer state.

 The ``purity'' of an RDM is a number which provides a sufficient condition to demonstrate that the RDM is dominated by a single state. 
The purity concept is widely used in quantum information theory \cite{Horodecki2009}. 
It can be used as both an entanglement and a decoherence assessment tool \cite{Bandyopadhyay2009}.
It is defined for the nuclear motion reduced density operator as 
\begin{eqnarray}
P=Tr[\hat{\rho}_\nuc^2]\ .
\end{eqnarray} 
We easily see that, when the molecular state is non degenerate, or when Eq.(\ref{elec-deg}) is satisfied, $P$ can take values between $1$, when a pointer state has probability one and all the others zero, and 
$\frac{1}{N_{dim}}$, when all pointer states are equiprobable. In case of a $n$-degenerate eigenstate leading to inequality (\ref{ens_ineq}), the maximum value for $P$ is $\frac{1}{n}$, and it is   achieved when each  $ \hat{\rho}_{\nuc}^i$ is dominated by a single pointer state.

Note that $P$ does not depend upon the basis set, so it can be evaluated
even if the pointer state basis has not been determined.
A value close to one implies that  one eigenvalue of  $\hat{\rho}_\nuc$ dominates all the others. The associated eigenstate can be considered as the dominant pointer state. 

In the BO approach, for example, the purity of $\hat{\rho}_\nuc$ is exactly $1$ in the non-degenerate case, since only the $\Psi_{N}$ appearing in Eq. (\ref{rdm_nuc_BO_el}) is populated. Such a state, at least the vibrational ground state of semi-rigid molecules (assuming separability of the rotational and translational DOFs), is well-localized in the neighbourhood of the so-called ``equilibrium geometry'' of the system. So, a molecular structure is recovered in this sense. 
However, it is often pointed out that recovering  molecular structure from the BO approach is not a great achievement, since it is put in from the start. 

\subsection{Environment classes defined by pointer states}

We have seen that, in decoherence theory, there is a set of pointer states (defined up to unitary transformations within equiprobable subsets) for every environment. 
Here we consider the inverse mapping. We select a set of orthonormal states, $\mathcal{S}$, of an isolated quantum system, and define the class, $\mathcal{E}$, of environments  for which 
we get the $\mathcal{S}$-set of pointer states. That is to say, the off-diagonal elements of the nuclear RDM in the set $\mathcal{S}$ representation,  $\hat{\rho}_\nuc$ (see Remark 2)  decays exponentially with time.  

In the following, we will extend this definition to a set, $\mathcal{S}$,  of (non-normalizable) Dirac distributions over the nuclear configuration space and assume that its associated environment class,  $\mathcal{E}$, is non empty. Note, however that $\mathcal{E}$ may contain more than a unique environment: for example, collisions by different types of particles may result in the same localization effect for a system in a bath of an appropriate colliding particle density.

\section{Decoherence by the electronic environment}

In this section, we consider a stationary eigenstate of the total system (electrons plus nuclei) Hamiltonian, and study the decoherence effect of the electrons on nuclear motion. The density operator of such an eigenstate, and therefore its reduced density matrices, being time independent, we do not aim at the determination of isomer lifetimes. 
 
We will assess the decoherence effect in two complementary ways. First, we will consider the electrons as the sole environment of the nuclei. In this context, the pointer states of interest for the molecular structure problem,  are the eigenstates of  $\hat{\rho}_\nuc$. They can be readily obtained and analysed. Second, we will study $\hat{\rho}_\nuc$ in a representation over a set of Dirac distributions in the nuclear configuration space. Such a set could constitute pointer states selected by an external environment that localizes the nuclei in space. We will assess the contribution of the electronic environment to the suppression of interferences between these potential pointer states.

Numerical examples are presented for the $H_2$ isotopologues,  using accurate electron-nucleus wave functions.
A complementary analysis, in which the nuclear reduced density matrix is analyzed for a broader range of systems, is presented in Ref.~\cite{MaCa20}.

\subsection{Purity of $\hat{\rho}_\nuc$ for $H_2$ isotopologues}
In this section, we consider a translation-free system and separate also the rotational DOFs 
in a way that only results in an effective, $J$-dependent term in the potential energy of the internal coordinate \cite{Cassam15-pra}.
Nuclear spin DOFs were not explicitly considered in our calculations, although they can prove important to take into account, 
when studying localization issues \cite{Bouakline2020}. However, they cannot be ignored in building degenerate eigenspaces fulfilling
the constraints imposed by Pauli spin-statistics theorem \cite{Pauli1940}. More precisely, for even $J$-values only nuclear spin singlets are allowed, so that the eigenspace is only  $(2J+1)^2$-degenerate,  whereas for odd $J$-values only nuclear spin triplets are allowed, so that the eigenspace is $3\times(2J+1)^2$-degenerate. To treat all  degenerate components on an equal footing, an ensemble density operator is necessary, as explained in section \ref{ens_state}, resulting in a sum of nuclear RDM satisfying relations (\ref{ens_orth}). However, we will consider that this operator has been further reduced by tracing over the orientational coordinates  (i.e. Euler angles) and nuclear spin coordinates. After this reduction, we reach a situation corresponding to Eq.(\ref{elec-deg}), since within our treatment of diatomic systems, the part depending upon Euler angles is factored out, leaving just an extra term in the internuclear potential which depends upon the angular momentum quantum number. So, maximal purity will be $1$ as in a non-degenerate case. The RDM becomes just the  operator, $\hat{\rho}_\nuc$, that is obtained by ignoring all nuclear coordinates except the internuclear distance. So the nuclear configuration $\mathbf{R}$ and $\mathbf{R}'$  appearing in  Eq.(\ref{rdm_nuc_el}) needs only to be specified by internuclear distances.

In Electron-Nucleus Full Configuration Interaction (EN-FCI) calculations \cite{Cassam15-pra,Cassam17-tca}
 a basis set of electronic states,  obtained at one and the same clamped nuclei configuration,  is used to build direct product, electron-nucleus basis sets. In Ref. \cite{Cassam15-pra,Cassam17-tca}, the basis set was not complete, but uses typically tens of thousands of electronic states for H$_2$, so many more than in Ref.\cite{Hu2018}. 
 Computing the vibrational reduced density matrix by tracing out the electronic degrees of freedom, or the electronic reduced density matrix by tracing out the vibrational degrees of freedom, we find that the approximate $(J=0)$-ground state basis function has a population of about $99\%$ for H$_2$, while the largest coherence between the approximate ground state and excited states basis functions are on the order of a few percents. For the second excited state, the approximate first excited vibrational basis function has a population of about $97\%$ for H$_2$. 
 
Let us focus now on the representation-free purity of the RDM for the vibrational DOF, $\hat{\rho}_\nuc$, (identical to that for the electrons by an extension of Carlson and Keller duality~\cite{Carlson61} since the electronic DOFs and the vibrational DOF are two complementary sets i.e. form a partition of the set of DOFs), is reported in Tab.\ref{tab:purity} for selected, molecular eigenstates, obtained with the calculation described in Ref.\cite{Cassam15-pra}. The purity values are sufficiently converged to address the changes with respect to rotational and vibrational excitation and isotopic substitution. The entries in italics correspond to \textit{a priori} less accurate, approximate molecular states. (Note that the label ``$^1\Sigma_g^+ 0\rightarrow 1$'' in Tab.~VI of \cite{Cassam15-pra} was referring to the transition  $X^1\Sigma_g^+ \rightarrow B^1\Sigma_u^+$).

\begin{table}[ht]
    \centering

\begin{tabular}{@{}llll@{}}
States&$\mathrm{H}_2$&$\mathrm{D}_2$&$\mathrm{T}_2$\\[5pt]
\hline
&&& \\[-5pt]
 $X^1\Sigma_g^+\ \nu=0,\ J=0$  & 0.988841  &  0.992048    &  0.993479  \\[5pt] 
 $X^1\Sigma_g^+\ \nu=0,\ J=1$  & 0.988830  &  0.992044    &  0.993476  \\[5pt]
 $X^1\Sigma_g^+\ \nu=0,\ J=2$  & 0.988807  &  0.992036    &  0.993472  \\[5pt]
 $X^1\Sigma_g^+\ \nu=0,\ J=3$  & 0.988773  &  0.992024    &  0.993466  \\[5pt]
 $X^1\Sigma_g^+\ \nu=0,\ J=4$  & 0.988728  &  0.992009    &  0.993458  \\[10pt]
 $X^1\Sigma_g^+\ \nu=1,\ J=0$  & 0.966507  &  0.976131    &  0.980428  \\[5pt]
 $X^1\Sigma_g^+\ \nu=1,\ J=1$  & 0.966473  &  0.976120    &  0.980420  \\[5pt]
 $X^1\Sigma_g^+\ \nu=1,\ J=2$  & 0.966406  &  0.976097    &  0.980409  \\[5pt]
 $X^1\Sigma_g^+\ \nu=1,\ J=3$  & 0.966306  &  0.976062    &  0.980390  \\[5pt]
 $X^1\Sigma_g^+\ \nu=1,\ J=4$  & 0.966173  &  0.976016    &  0.980365  \\[10pt]
 $X^1\Sigma_g^+\ \nu=2,\ J=0$  & \textit{0.944993}  &  0.960590    &  0.967589  \\[5pt]
 $X^1\Sigma_g^+\ \nu=2,\ J=1$  & \textit{0.944935}  &  0.960571    &  0.967579  \\[5pt]
 $X^1\Sigma_g^+\ \nu=2,\ J=2$  & \textit{0.944822}  &  0.960535    &  0.967558  \\[5pt]
 $X^1\Sigma_g^+\ \nu=2,\ J=3$  & \textit{0.944655}  &  0.960480    &  0.967527  \\[5pt]
 $X^1\Sigma_g^+\ \nu=2,\ J=4$  & \textit{0.944440}  &  0.960408    &  0.967486  \\[10pt]
 $B^1\Sigma_u^+\ \nu=0,\ J=0$  & 0.983095  &  \textit{0.987301}    &  \textit{0.989288}  \\[5pt]
 $B^1\Sigma_u^+\ \nu=0,\ J=1$  & 0.983104  &  \textit{0.987308}    &  \textit{0.989119}  \\[10pt]
 \hline
\end{tabular}
   \caption{{Purity of $\hat{\rho}_\nuc$ for selected electron-nucleus eigenstates. The convergence of the values in italics has not been well established, and they are not used in making any conclusion in the text. $J$ is the rotational angular momentum quantum number, $\nu$ an approximate vibrational quantum number, and the first label of each row designates the approximate label for the electronic state.}}
    \label{tab:purity}
\end{table}

 The first observation is that all purity numbers are close, but not equal, to one, so that all the corresponding molecular eigenstates are reasonably, but not perfectly pure. There are two clear tendencies. Firstly, following every row, we note that purity numbers increase. This is not surprising: it is related to the decrease of the De Broglie wave length with increasing nuclear mass, and the corresponding increase in state localization, illustrated in Fig.~3 of Ref.\cite{Cassam17-tca}. Secondly,  across every column,
purity tends to decrease  with vibrational and rotational excitation for an electronic state, and for successive  approximate electronic states  of a given approximate  ro-vibrational state. For example, for H$_2$, the purity is  $0.983$ for the lowest ro-vibrational states of the first singlet, electronic, excited state, $B^1\Sigma_u^+\ \nu=0,\ J=0$ and $J=1$, while it is $0.989$ for states $X^1\Sigma_g^+\ \nu=0,\ J=0$ and $J=1$. 

\begin{figure}[ht]
    \centering
    \includegraphics[scale=0.4,angle=90]{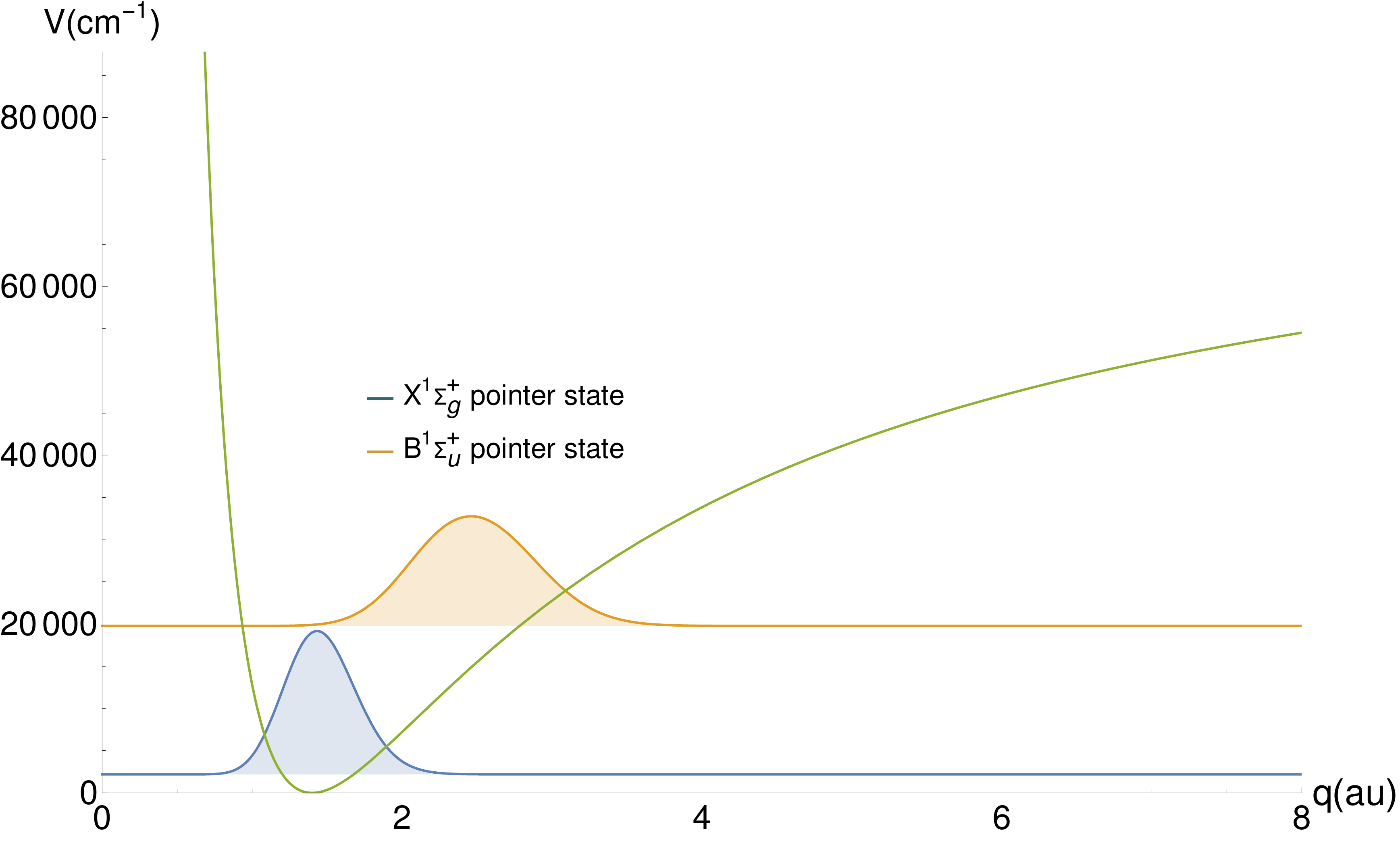}
    \caption{Dominant pointer states (in arbitrary units) of the two lowest ``electronic'' singlet states of H$_2$ as a function of the internuclear distance $q$ (in Bohr). The Kratzer potential curve \cite{Cassam15-pra}, $V$ in \cm, of the vibrational Hamiltonian whose lowest eigenfunctions were used as the nuclear motion basis set, is also shown. The vertical alignment of the pointer states with respect to the minimum of the potential curve and with respect to each other, is arbitrary. }
    \label{fig:pointer}
\end{figure}

The purity values from our $\hat{\rho}_\nuc$ are quite different to what is found in the simple dynamical model, reduced to the first two electronic states of Ref.\cite{Hu2018}. 
The purity of $\hat{\rho}_\nuc$ for low-lying molecular eigenstates is always found to be high. So, the outcome of nuclear position measurements will be distributed according to the module square of the most populated eigenstate of $\hat{\rho}_\nuc$ with high probability. The larger the mass of the system, the larger the probability and the localization of the density. So, the pointer states for the electronic environment alone, that is to say, the eigenstates of $\hat{\rho}_\nuc$, will tend to Dirac distributions only at the infinite mass limit. The $(J=0)$-pointer states of the main isotopologue are depicted in Fig.\ref{fig:pointer}. They are very similar to the vibrational eigenstates  (not shown in the Figure) one would obtain in the BO approach.

The analysis can be generalized to polyatomic molecules. One can obtain a nuclear motion RDM by tracing out the electronic DOFs.
In molecular systems with multiple large-amplitude motions, it is not always possible to separate out the rotational motion from the other nuclear degrees of freedom \cite{Schmiedt15}.  The present discussion does not consider these special systems for which a classical-like structure is hardly relevant.

To conclude this section, the purity of the nuclear reduced density matrix is a representation-free measure that allows to determine whether the electron-nucleus wave function is dominated by a single  state for nuclear internal DOFs. If the purity is close to one, as we saw for the example of molecular hydrogen (and isotopologues),
then, the nuclear structure is well characterized by the nuclear density of the dominant state. If the latter density has a single, narrow maximum, then a ``classical-like'' molecular structure emerges, since according to Born's probabilistic interpretation, repeatedly probing the nuclei relative positions, will consistently find values around the sharp maximum.
So, the electronic environment seems able to explain at least partially the classical-like internal structure of semi-rigid molecules.

However, such an approach cannot explain the non observation of the superposition of parity-broken enantiomers, nor the breaking of molecular orientational symmetry, because a  system of electrons plus nuclei having only a finite number of DOFs cannot partition the Hilbert space into superselection sectors \cite{Wick1952,Haag1964}, as a consequence of the Stone-Von Neumann theorem \cite{Amann88,Strocchi88}. A natural explanation to limit the dynamical instability due to superpositions, and so to induce effective superselection rules, is to introduce an external environment such as in Ref. \cite{Trost2009} for parity-breaking in chiral molecules or Ref. \cite{Stickler2018a} for orientational symmetry-breaking.
In the next section, we stay in the framework of an isolated molecular system, however, we study the RDM in a representation of 
plausible, localized states in the nuclear configuration space that can be selected as pointer states by an external environment. 

\subsection{Interference damping of rotational pointer states}

So, let us consider a set of perfectly localized states, $\mathcal{S}$. Assuming that their associated environment class, $\mathcal{E}$, is non-empty, they could constitute a set of pointer states for our system, that is to say, the RDM (reduced for electron plus environment DOFs, see remark 3) would be diagonal when represented in this set. In contrast, the RDM (reduced for electron DOFs only) of the isolated molecule is expected to have non-zero non-diagonal elements. The deviation from zero of the latter will tell us, how far the elements of set $\mathcal{S}$ are from the pointer state status.

To evaluate in this way the contribution of the electronic subsystem to nuclear position decoherence, we take the same example of
$\mathrm{H}_2$ isotopologues.
The study will be limited to  the ground rovibronic eigenstate with zero total angular momentum of the isolated system. Its reduced density operator $\hat{\rho}_\nuc$ will be further reduced to the one-nucleus density, $\hat{\rho}_{0,n}$, with respect to the center of mass \cite{LuEcLoUg12,BePoLu13}.

\begin{figure}
\includegraphics[scale=1.0]{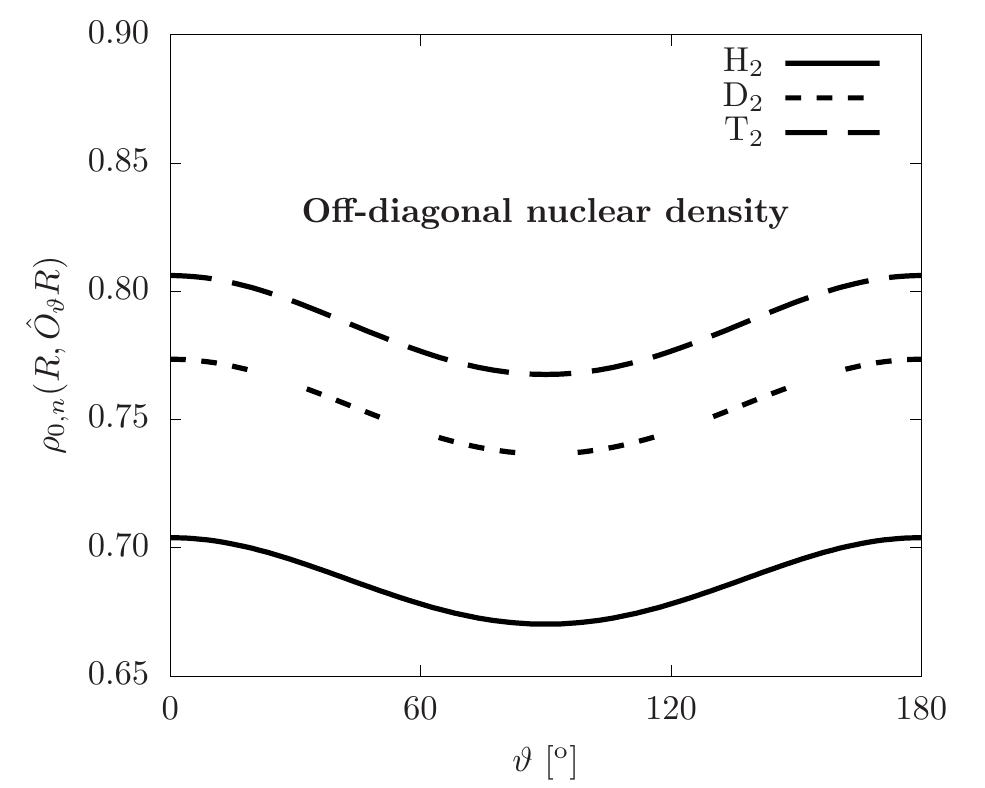}
  \caption{
    Off-diagonal nuclear density at the internuclear distance of maximal diagonal density for $\mathrm{H}_2$ isotopologues.  
    The curves are interpolated from a grid of 100 points regularly spaced.
    \label{fig:H2like}
  }
\end{figure}

The molecular wave function is spherically symmetric, and only the relative angular difference between two pointer states matters.
Then, fixing the internuclear distance between the two nuclei and a plane, $\mathcal{P}$, containing them,  allows us to restrict the set, $\mathcal{S}$ to a set of configuration-centered delta distributions corresponding to nuclear positions rotated around the center-of-mass within $\mathcal{P}$. Each of these states can be specified by a single angle parameter, $\vartheta$. 
Assigning the zero angle to some reference position, state  $|\xi_\vartheta\rangle\in \mathcal{S}$ will be related to $|\xi_0\rangle$, the state corresponding to this reference nuclear configuration, by $|\xi_\vartheta\rangle=|\hat{O}_\vartheta\xi_0\rangle$ where $\hat{O}_\vartheta$ is the rotation operator  in plane $\mathcal{P}$ of angle $\vartheta$ around the center of mass. 
We study
orientational decoherence  due to the electrons  by calculating the damping of cross-terms between  these states, $\rho_{0,n}(R,\hat{O}_\vartheta R):=\langle\xi_0|\hat{\rho}_{0,n}|\xi_\vartheta\rangle$, as a function of their angular distance, see Fig.\ref{fig:H2like}.

The ground-state wave function of the four-particle systems was computed
using an explicitly correlated Gaussian basis set and the QUANTEN computer program \cite{Ma19review} (see also Refs.~\cite{MaHuMuRe11a,MaHuMuRe11b} relevant for this work).
The aim was to get RDM matrix elements  for
the ground state of these system with zero total angular momentum ($N=0$), natural parity
($p=+1$), and zero spin for the pair of electrons, and of positive particles. We managed to converge the corresponding energies within 1~$\%$  and we have checked that it was sufficient to obtain  converged curves for the different $\mathrm{H}_2$ isotopologues within the resolution of Fig.\ref{fig:H2like}, where the one-nucleus reduced density matrix elements,
$\rho_{0,n}(R,\hat{O}_\vartheta R):=\langle\xi_0|\hat{\rho}_{0,n}|\xi_\vartheta\rangle$  are displayed.  For the sake of simplicity, we have fixed plane $\mathcal{P}$ to define the angle $\vartheta$, and the pointer states $|\xi_\vartheta\rangle$. However, thank to the spherical symmetry, we may as well consider that the coordinates of the two nuclei
are at antipodal points of a sphere centered around their midpoint. 

To understand when the interference terms get small and the localization of the nuclei by the electrons efficient,
it is convenient to return to the Born--Oppenheimer approximation, Eq.~(\ref{rdm_nuc_BO_el}).
If the overlap of the electronic wave function corresponding to the rotated nuclear structures
is small, then $\langle\xi_0|\hat{\rho}^{\text{[BO]}}_{0,n}|\xi_\vartheta\rangle$ is also small.
More generally, interferences are damped if the electronic cloud of the molecular wave function 
changes significantly between the rotated nuclear configurations. 

The results of Fig.\ref{fig:H2like}, shows that for rotated H$_2$ isotopologues, the contribution of electrons to interference suppression is 
small, with only about 4~\% suppression at $90^{\circ}$. All the curves are parallel, showing that the relative suppression effect with respect to   $\langle\xi_0|\hat{\rho}_{0,n}|\xi_0\rangle$ is almost mass-independent.
In a more complete study~\cite{MaCa20}, 
it has been found that the lighter Ps$_2=\{\text{e}^+,\text{e}^+,\text{e}^-,\text{e}^-\}$ system
 retains full coherence with respect
to orientational changes with less than 1~\% suppression at most.
Conversely, one can expect a larger suppression effect for heavier atoms with a strong electronic core density, since the overlap between the electronic part of the wave function and that of the rotated structure would be smaller.

\section{Conclusion}

The concept of molecular structure with fixed values of geometrical parameters, is at odds with quantum mechanics because of the uncertainty principle. Furthermore, in quantum mechanics, we can only ask what is the structure of the molecule in a given molecular state. Certainly, the average geometrical parameters can change drastically upon molecular excitation. For example, the ground-state equilibrium geometry of HeH$^+$ is $\approx 1.463$~bohr~\cite{Kolos76a}, while it is $\approx 5.53$~bohr in its first excited $^1\Sigma^+$ electronic state~\cite{Kolos76b}.

%
Then, one could argue that in the fundamental theory of elementary particles interacting by electromagnetic interactions (quantum electrodynamics, QED), excited states are not stable due to spontaneous emission, so
the ground state has a distinctive role. 
In the present work, we addressed the molecular structure problem within the standard non-relativistic quantum mechanics but without invoking the usual BO (nor the stritcly speaking ``adiabatic'') approximation. The molecule was described with an electron-nucleus wave function and we focused on (semi-rigid) molecular systems for which a molecular structure can be attributed (in their ground or lowest excited states) within the BO theory. That is to say, the scope of the present study is restricted to molecules that appear classical-like to chemists, (even if the finite-temperature chemistry that has forged their intuition would require thermally averaged ensemble density operators, rather than those representing single Hamiltonian eigenstates employed in our study).

We have considered the electrons as an environment continuously  probing the positions of the nuclei.
From obvious electrostatic considerations, nuclei without electrons would not be bound, their wave functions would be plane waves, and no structure would emerge.
We have used the concept of purity of the nuclear reduced density matrix, that is a representation-free measure, to analyse the pointer states, selected by the electronic environment alone, for the internal structure of the nuclei.
Numerical examples were presented for the H$_2$ molecule and its isotopologues within electron-nucleus computations (without relying on the BO nor adiabatic approximations).

Every studied low-lying state of the dihydrogen molecule was dominated by a single pointer state that has similar
characteristics to the vibrational state of the BO theory, but these nuclear pointer states were obtained as eigenfunctions of the nuclear reduced density matrix computed from the electron-nucleus wave function. The procedure can, in principle, be generalized to the electron-nucleus wavefunction of polyatomic molecules, and provides a route to obtain nuclear pointer states, and derive a number of features from them, including structural information, without invoking the BO nor adiabatic approximations.

Then, we have quantified the suppression of the non-diagonal density matrix elements due to the electronic part of the wave function, between perfectly localized states. We have found only a small suppression effect on the H$_2$ molecule and its homonuclear isotopologues. However, for molecules that include nuclei with higher nuclear charges, we may expect that the stronger electron-nucleus Coulomb interaction would increase the damping of interference terms.

\section*{Acknowledgements}
This project was initiated within a Short Term Scientific Mission of the MOLIM COST Action. 
EM acknowledges financial support from a PROMYS Grant (no. IZ11Z0 166525) 
of the Swiss National Science Foundation.

\end{document}